\documentclass{appolb}
\usepackage{graphicx}

\newcommand{\tso}{${}^{3\!}S_1$}
\newcommand{\tdo}{${}^{3\!}D_1$}

\begin{document}
\title{Unquenching the meson spectrum: \\ a model study of exited
$\rho$ resonances\thanks{Talk by G.~Rupp at Workshop ``Excited QCD 2016'',
Costa da Caparica, Portugal, March 6--12, 2016.}
}
\author{George Rupp\address{Centro de F\'{\i}sica e Engenharia de Materiais
Avan\c{c}ados, Instituto Superior T\'{e}cnico, Universidade de Lisboa,
P-1049-001, Portugal}
\\[5mm]
Susana Coito\address{Institute of Modern Physics, CAS, Lanzhou 730000, China}
\\[5mm]
Eef van Beveren\address{Centro de F\'{\i}sica da UC, Departamento de
F\'{\i}sica, Universidade de Coimbra, P-3004-516, Portugal}
}
\maketitle

\begin{abstract}
Quark models taking into account the dynamical effects of hadronic decay
often produce very different predictions for mass shifts in the hadron
spectrum. The consequences for meson spectroscopy can be dramatic and 
completely obscure the underlying confining force. Recent unquenched
lattice calculations of mesonic resonances that also include meson-meson 
interpolators provide a touchstone for such models, despite the present
limitations in applicability. On the experimental side, the $\rho(770)$ 
meson and its several observed radial recurrences are a fertile testing
ground for both quark models and lattice computations. Here we apply
a unitarised quark model that has been successful in the description of many
enigmatic mesons to these vector $\rho$ resonances and the corresponding
$P$-wave $\pi\pi$ phase shifts. This work is in progress, with encouraging 
preliminary results.
\end{abstract}

\section{Introduction}
The static quark model, which describes hadrons as pure bound states of
confined quarks and antiquarks, has remained largely unchallenged for about
40 years. Even nowadays most experimentalists still confront the
enhancements in their meson data with the relativised quark model of
Godfrey and Isgur (GI) \cite{PRD32p189} in order to arrive at an assignment
or otherwise claim to have found evidence of some exotic state. Now, the
GI model is indeed the most comprehensive calculation of practically all
possible quark-antiquark masses, employing the usual Coulomb-plus-linear
(``funnel'') confining potential. However, the insistence on comparing both
narrow and broad structures in cross sections directly with the infinitely
sharp levels of a manifestly discrete confinement spectrum is clearly a
poor-man's approach. Yet, models going beyond the static quark model
have been around for almost the same four decades, the pioneering ones
being the Cornell model for charmonium \cite{PRD17p3090}, the Helsinki
model for light pseudoscalars and vectors \cite{AOP123p1,ZPC5p205}, and
the Nijmegen model for heavy quarkonia \cite{PRD21p772} and all
pseudoscalar \& vector mesons \cite{PRD27p1527}. Despite the at times huge
mass shifts predicted by these models, for many years the effects of
decay, also called coupled-channel contributions or unitarisation, were
largely ignored.  Instead, inspired by perturbative QCD, hadron
spectroscopists made their models more and more sophisticated at the level
of the confining potential, with e.g.\ spin-orbit splittings and also
relativistic corrections \cite{PRD32p189}, which are nevertheless quite
insignificant as compared to many of the large mass shifts from unitarisation.
\\ \indent
Only since the observation of a growing number of enigmatic mesons, whose
masses or observed decays do not seem to fit in the GI and similar static
quark models, more authors started to take into account
dynamical effects from strong decay and scattering. Parallelly, very
recent unquenched lattice computations have shown remarkably large mass
shifts due to the inclusion of two-meson interpolators besides the usual
quark-antiquark ones, thus confirming the importance of decay for meson
spectroscopy. \\ \indent
An appropriate class of mesons to study these issues is $\rho(770)$
and its several radial excitations, together with the corresponding
$P$-wave $\pi\pi$ phase shifts, because of the considerable amount of
available data, despite being mostly old  \cite{PDG2014}. Here we shall
present preliminary results in the context of the Resonance-Spectrum
Expansion (RSE), which is a momentum-space variant of the unitarised model
employed in Ref.~\cite{PRD21p772}.  \\ \indent
In Sec.\ 2 meson mass shifts in different quark models that include hadronic
decay are compared, also with a recent lattice calculation. Section~3 is
devoted to a very brief description of the RSE model as applied to the
isovector vector mesons, with some preliminary yet encouraging results. A
few conclusions are drawn in Sec.~4. 
\section{Mass shifts from ``unquenching'' in models and on the lattice}
Quark models that dynamically account for decay are often called ``unquenched''
\cite{JPG31p845,AIP947p168,IJMPCS02p178,PRD90p034009,EPJC75p26}. Now, this is
actually a very sloppy name, as the term ``unquenched'' originates in lattice
calculations with dynamical instead of static quarks, via a fermion
determinant. We shall nevertheless use this inaccurate name when referring
to such quark models, because the various approaches are very different. For
instance, Refs.~\cite{AIP947p168} and \cite{PRD90p034009} evaluate real or
complex mass shifts from lowest-order hadronic loops, Ref.~\cite{IJMPCS02p178}
constructs and uses a screened confining potential supposedly
resulting from quark loops, while Ref.~\cite{EPJC75p26} includes meson loops to
all orders in a fully unitary $S$-matrix formalism. Also the original models
of Refs.~\cite{PRD17p3090,PRD21p772,PRD27p1527} were truly unitarised. But
there are enormous differences as well in the computed mass shifts from
unquenching, even among in principle similar models. In Table~\ref{massshifts}
 \begin{table}[h]
\caption{Negative mass shifts from unquenching. Abbreviations: BT = bootstrap,
$\chi$ = chiral, QM = quark model, RGM = Resonating Group Method, RSE =
Resonance Spectrum Expansion, CC = coupled channels, HO = harmonic oscillator,
WF = wave function, PT = perturbation theory, $q$ = light quark; $P,V,S$ =
pseudoscalar, vector, scalar meson, respectively.}
\begin{center}
\begin{tabular}{l|l|l|c}
\hline\hline
Refs.\ & Approach &  Mesons & $-\Delta M$ (MeV) \\ \hline
\cite{PRD17p3090} & $S$-matrix, $r$-space & charmonium &
48--180  \\
\cite{AOP123p1,ZPC5p205} & one-loop BT & light $P,V$ & 530--780, 320--500 \\
\cite{PRD21p772,PRD27p1527} & $S$-matrix, $r$-space & 
$q\bar{q}$, $c\bar{q}$, $c\bar{s}$, $c\bar{c}$, $b\bar{b}$; $P,V$ &
$\approx\!$ 30--350 \\
\cite{ZPC30p615} & $S$-matrix, $r$-space & 
light, intermediate $S$ & 510--830, $\sim0$\\
\cite{PRD42p1635} & $\chi$ QM, RGM & 
$\rho(770)$, $\phi(1020)$ & 328, 94\\
\cite{PRL91p012003}  & RSE, $p$-space &
$D_{s0}^\star(2317)$, $D_0^\star(2400)$ & 260, 410 \\
\cite{PRD70p114013} & CC, $\chi$ Lagrangian &
$D_{s0}^\star(2317)$, $D_s^\star(2632)$ & 173, 51 \\ 
\cite{PRD72p034010} & CC, HO WF & charmonium & 165--228 \\ 
\cite{PRC77p055206} & CC, PT & charmonium & 416--521 \\
\cite{EPJC71p1762} & RSE, $p$-space & $X(3872)$ & $\approx\!100$ \\
\cite{PRD84p094020}  & RSE, $p$-space &  
$c\bar{q}$, $c\bar{s}$; $J^P=1^+$ & 4--13, 5--93  \\ \hline
\end{tabular}
\end{center}
\label{massshifts}
\end{table}
we show the corresponding predictions of a number of unquenched quark models
for mesons.  Note that the mass shifts in Refs.\
\cite{PRD21p772,PRD27p1527,ZPC30p615,PRL91p012003,EPJC71p1762,PRD84p094020} are
in general complex, in some cases \cite{ZPC30p615,PRL91p012003,PRD84p094020}
with huge imaginary parts, corresponding to pole positions in an exactly solved
$S$-matrix. As for the disparate shifts among the various approaches, they are
due to differences in the assumed decay mechanism, included channels, and
possibly drastic approximations. Another crucial point should be to properly
account for the nodal structure of the bare $q\bar{q}$ wave functions.
\\ \indent
Faced with these discrepancies, one is led to look at unitarised lattice
results, preferably on $\rho(770)$ and its radial recurrences,  which we
will study here with the RSE formalism. Unfortunately, no such calculations
have been published so far. Nevertheless, a recent paper \cite{PRD88p054508} on
the related vector meson $K^\star(892)$ and the associated $P$-wave $K\pi$
phase shifts provides very useful information. Not only were the $K^\star$ mass
and extrapolated width reasonably well reproduced, but also a prediction,
albeit approximate, was made for the first radial excitation $K^\star(1410)$
\cite{PDG2014}, finding a mass of $1.33\pm0.02$~GeV. Now, the main surprise
about the latter number is not so much its relative closeness to the
experimental value and the 250-MeV gap with e.g.\ the ``quenched'' GI
\cite{PRD32p189} prediction. Rather, being about 300 MeV lower than the value
found by the same lattice group in another unquenched calculation
\cite{ARXIV13116579} yet with no two-meson interpolators included, it showcases
the potentially dramatic effects of unitarisation on meson spectra. This is an
excellent incentive to study the $\rho$ spectrum and $P$-wave $\pi\pi$ phases
in detail.
\section{RSE modelling of \boldmath{$\rho$} recurrences and \boldmath{$P$}-wave
\boldmath{$\pi\pi$} scattering}
The experimental status of radial $\rho$ excitations was reviewed minutely in
Ref.~\cite{ARXIV151000938}. Suffice it here to stress the clearly biased
handling of a frequently reported $\rho(1250)$ resonance by the Particle Data
Group (PDG), by lumping some of its observations under $\rho(1450)$ 
\cite{PDG2014}, instead of creating a separate entry in the meson listings. The
PDG also bluntly omits a reference to a relatively recent phase-shift anaysis
\cite{NPA807p145} that concludes $\rho(1250)$ to be the most important
$\rho$ excitation in order to fit the data. Moreover, there is the
well-established excited $s\bar{q}$ resonance $K^\star(1410)$ (see PDG
\cite{PDG2014} summary table), now also confirmed on the lattice
\cite{PRD88p054508}. This lends further evidence to the existence of 
$\rho(1250)$, as predicted long ago in the model of Ref.~\cite{PRD27p1527}.
 
The general expressions for the RSE off-energy-shell $\mathcal{T}$-matrix and
corresponding on-shell $\mathcal{S}$-matrix have been given in several
papers (see e.g.\ Ref.~\cite{PRD84p094020}). In the present case of $P$-wave
$\pi\pi$ scattering, the quantum numbers of the system are
$I^GJ^{PC}=1^+\,1^{--}$, which couples to the $I\!=\!1$ quark-antiquark state
$(u\bar{u}-d\bar{d})/\sqrt{2}$ in the spectroscopic channels \tso\ and \tdo.
In the meson-meson sector, we only consider channels allowed by total angular
momentum $J$, isospin $I$, parity $P$, and when possible G-parity $G$. The
included combinations from the lowest-lying meson nonets \cite{PDG2014} are:
PP, VP, VV, VS, AP, and AV, where P stands for $J^{PC}=0^{-(+)}$, V for
$1^{-(-)}$, S for $0^{+(+)}$, and A for $1^{+(+)}$ or $1^{+(-)}$. 
This choice of meson-meson channels is motivated by the
observed two- and multi-particle decays of the $\rho$ recurrences up to 
$\rho(1900)$ \cite{PDG2014}, which include several intermediate states
containing resonances from the referred nonets. For instance, the PDG lists
\cite{PDG2014} under the $4\pi$ decays of $\rho(1450)$ the modes
$\omega\pi$, $a_1(1260)\pi$, $h_1(1170)\pi$, $\pi(1300)\pi$, $\rho\rho$, and
$\rho(\pi\pi)_{\mbox{\scriptsize $S$-wave}}$, where
$(\pi\pi)_{\mbox{\scriptsize $S$-wave}}$ is probably dominated by the
$f_0(500)$ \cite{PDG2014} scalar resonance. By the same token, the $6\pi$
decays of $\rho(1900)$ will most likely include important contributions
from modes as $b_1(1235)\rho$, $a_1(1260)\omega$, etc.. For consistency of
our calculation, we generally include complete nonets in the allowed decays,
and not just individual modes observed in experiment. The only exception is the
important $\pi(1300)\pi$ P$^\prime$P mode, because no complete nonet of
radially excited pseudoscalar mesons has been observed so far \cite{PDG2014}.
The resulting 26 channels are given in Table~\ref{channels}. \\ \indent
\begin{table}[!h]
\caption{Included classes of decay channels. Short-hand: $\sigma$ = $f_0(500)$,
$a_0$ = $a_0(980)$, $\kappa$ = $K_0^\star(800)$, $a_1$ = $a_1(1260)$,
$b_1$ = $b_1(1235)$, $h_1$ = $h_1(1170)$, $K_1$ = $K_1(1270)$,
$\tilde{K}_1$ = $K_1(1400)$, $f_1$ = $f_1(1285)$, $\pi^\prime$ = $\pi(1300)$
\cite{PDG2014}.}
\begin{center}
\begin{tabular}{c|l|c}
\hline\hline
Nonets & \hspace*{9mm}Two-Meson Channels &  $L$ \\ \hline
PP & $\pi\pi$, $KK$ & 1 \\ 
VP & $\omega\pi$, $\rho\eta$, $\rho\eta^\prime$, $K^\star K$ & 1 \\ 
VV & $\rho\rho$, $K^\star K^\star$  & 1 \\ 
VS & $\rho \sigma$, $\omega a_0$, $K^\star\kappa$ & $0,2$ \\
AP & $a_1\pi$, $b_1\eta$, $b_1\eta^\prime$,
     $h_1\pi$, $K_1K$, $\tilde{K}_1K$ & 0 \\ 
AV & $a_1\omega$, $b_1\rho$, $f_1\rho$,
     $K_1K^\star$, $\tilde{K}_1K^\star$ & 0 \\ 
P$^\prime$P & $\pi^\prime\pi$ & 1  \\ \hline
\end{tabular}
\end{center}
\label{channels}
\end{table}
With the few available parameters \cite{ARXIV151000938}, a good fit to the
$P$-wave phase shifts is only possible up to about 1.2~GeV, whereabove the
phases rise a bit too fast, though their qualitative behaviour can be
reproduced. Improvements may require more flexibility in the transition
potential, by allowing different decay radii for the various classes of
two-meson channels, and/or allowing for complex-mass resonances in the
final states \cite{ARXIV151000938}. The present fit yields a reasonable
$\rho(770)$ pole, viz.\ at $(754-i67)$~MeV, while there are two poles
in the range 1.2--1.5~GeV, compatible with both $\rho(1250)$ and $\rho(1450)$.
\section{Conclusions}
Meson spectroscopists are slowly starting to leave the stone-age behind, by
realising that effects from strong decay can be of the same order as the
bare $q\bar{q}$ level splittings themselves. Enormous obstacles lie on the
road ahead, demanding more theoretical work, improved lattice calculations,
and much better experimental analyses. The excited $\rho$ spectrum provides
an excellent laboratory for such efforts. 
To make life even harder, several bumps in meson production
processes \cite{AOP323p1215} may just be non-resonant threshold enhancements
(see talk by E.~van Beveren \cite{ARXIV160503437}).

\end{document}